# A Decision Support System Web—Application for the Management of Forest Road Network

Apostolos Kantartzis[1] and Chrisovalantis Malesios[2]

1. Department of Forestry and Management of the Environment and Natural Resources, Democritus University of Thrace, Orestiada, Greece, 68200.

2. Department of Agricultural Development, Democritus University of Thrace, Orestiada, Greece, 68200.

**Abstract:** The present study contributes to the development of an online FRMP (Forest Road Management Platform) that aims to assist in the management of forest road network in a holistic way. This is achieved by the proposed methodology which serves as a database using geoprocessing and geospatial technologies for the handling, and the identification of critical issues in the infrastructure of forest road networks, visualization of forest roads, and the optimization of the management of the forest road network by proposing alternative strategies. In this paper, the development of the decision making web-tool, and present examples to demonstrate effectively its application and resulting advantages are described. The developed web-application may provide assistance to various forest organizations in the management of forest road networks and associated problems in an effective and sustainable way.

## 1. Introduction

The implementation of planning in the context of road network maintenance activities appears to be a necessity which significantly affects the policy of technical interventions at national, regional and local levels in each country. Until recently, each maintenance policy was distinguished from interventions whose main objective was to repair major operational malfunctions on low-quality pavements in terms of qualitative characteristics. However, contemporary perceptions on this issue go into a more in-depth analysis of the problem and propose a systematic information policy that is structured around three main axes: diagnosis-forecast-planning.

Nowadays, the need for an even more integrated management approach of woodlands as well as the flexibility that a full road network offers against the prevention of risk of forest fires have led forest authorities to the completion of the opening of many woodlands. Smooth access to all points of the forest road network throughout the year is a challenge as well as its proper management and organization for its maintenance. Building a unified database that will function as a platform for development within a forestry area can be considered to be the cornerstone for the generation of a developmental plan. Such a plan will include the registration and production of an electronic database that consists of the basic elements and of the main characteristics of the forest road network and the forest techniques applied so far.

One is often faced with contradictory situations such as addressing and managing forest road network problems that their full dimension is unknown. Thus, it

**Corresponding author:** Chrisovalantis Malesios, PhD, Marie-Curie Research Fellow, main research fields: applied statistics, environment and agriculture.

is evident that part of the solution to a problem is to record the knowledge of the quantity and quality of its data and then make decisions to adequately deal with it.

Forestry works are basic infrastructure projects for the development of the mountainous and national economy in general, and thus one needs to address them in conjunction with other problems associated with the protection and preservation of forest wealth.

The mountain inhabitants' economy is based on forests, which constitute a valuable resource. The need for further exploitation is evident, both for the improvement of the living standards of forestry and suburban population, as well as for the greatest possible tourist, aesthetic, hygienic and above all, their protective performance. However, in order to achieve and implement any development action in the mountainous area and especially in the forest areas, it is important to make necessary infrastructure works and in particular to develop the necessary network of forest roads.

Forest roads are an essential yet costly part of forest management, hence optimization methods are significant tools in the planning of road systems [1-2]. Forest road planning plays an important role in forest management and logging practices [3]. The importance of forest roads is stressed in various studies in the literature for a variety of reasons such as wood production and ecotourism [3], sustainability and environmental impact on forests [4].

On the other hand, forests roads are often considered to be controversial investments due to the fact that they may trigger or increase soil erosion [5-6], habitat fragmentation, [7] or illegal logging [8]. Hence the planning processes can become quite complex [9], because all these negative side-effects may grow with the density of the forest road network.

Thus, it is evident that forest road network planning plays a crucial role in fulfilling the goals of sustainable forest management. There are research studies on forest road planning [10-12].

Forest road network planning and maintenance therefore require the development of rational approaches to collecting and evaluating the technical, circulatory, economic data of each problem and to come up with concrete proposals for the required interventions taking into account all technological developments in the field of road construction. Towards these approaches, Dragoi, M., et al. [13] utilized benefit, cost and risk analysis on extending the forest roads network using a case study in Romania. It is evident from the above presentations of forest road management applications, that forest road network planning and management plays a crucial role in fulfilling the goals of a sustainable forest management system [10]. To the best of the authors' knowledge, there are no globally-adoptable tools and platforms for the decision making that may assist in the management of forest road network in a holistic way.

The aim of the current paper, is the presentation of an online FRMP (Forest Road Management Platform) that may assist in the management of forest road network in a unified way. The proposed computer application is aiming to adequately identify and handle potential critical issues in the infrastructure of forest road networks, visualization of forest roads, and the optimization of the management of the forest road network.

There are many applications that the proposed decision making web-tool can be utilized for. For example, the developed tool could be used to optimize the operation of the forest offices in relation to the forest road network (these may include marking trees for felling, transit of truck drivers to pick up timber and moving workers for logging), or members of other public services moving to a forest road network, such as directorate of coordination and supervision of forests or fire department, forest guard/game guard body or forest officers. Additionally, it can be exploited by individuals who visit and wish to approach remote parts of a forest, e.g. tourists, hunters.

It is believed that a modern view on this issue will offer a deeper analysis of the problem and will help suggesting a policy of maintenance on a systematic basis that will be organized around the already mentioned three pillars of diagnosis-forecast-programming [14].

The fully-integrated systems of maintenance usually consist of mathematical models that are equivalent to the three pillars mentioned above and they are used to manage and assess the data of a problem, from technical, circulatory and financial perspective, leading to specific solutions necessary for the interventions that have to be made defining at the same time the time needed for the execution, the size, type and cost of the operations. Such a system was applied in a specific area of investigation, i.e. the forest of Koupa, Thessaloniki, Greece.

Specifically, an integrated system consisting of various open source technologies and code for managing, in a (cost)-effective way, forest road networks has been designed.

Some of these technologies, developed in the current study are:

• An Online Forest Management Platform;
• An SQL Server to maintain and disseminate the data;
• A MapServer for serving of the spatial data.

The rest of the paper is organized as follows. Section 2 describes in detail the development of the proposed decision making web-tool for the management of forest road network system. Section 3 illustrates the applicability and advantages of the proposed methodology through the use of real-data examples. Finally, section 4 concludes with the main objectives and proposals of the current research.

## 2. Material and Methods

The systematic approach to forest road management and maintenance stems from the need for precise information on the actual situation of the road and road construction problems in order to predict their future situation and to compare and select between different route alternatives [15]. In this section, a detailed description of the development and application of the web-based forest road management tool that includes the recording and creation of a digital database on the characteristics of the forest road network and its forestry projects and proposes practical and reliable solutions will be presented.

The main objective is the exploitation of all data resulting from the use of these modern technologies (Orthophotos, orthophotographic maps, digital terrain models, land use cartographic polygons etc.) for the development of different forest applications using GISs (Geographic Information Systems).

The main feature of GIS is that they allow the connection between qualitative and descriptive features with spatial information. They are not simply a means of producing maps, charts or lists of qualitative features, but a new integrated technology necessary for the analysis and study of space and a decision making tool relating to the environment and man [16]. GISs were used using the national geographic road database in Sweden to take account of various factors (e.g. logging, spinning, costs, accessibility, etc.) to provide efficient planning for improving the forest road network to ensure a smooth supply of forest products to the corresponding forest industries [16]. Gumus, S., et al. [17] use GIS to assess the forest road network environmentally by looking at figures from various environmental indicators.

The structure of the application proposed in the current study, allows one to combine visual, geospatial data together with descriptors so that the user of the application can have the best possible understanding of the forest road network problem that has been reported. Thus, in overall, the proposed application can include a description of the problem, the location of the problem illustrated on a forest map in order to be able to estimate with the help of the program the severity, cost and the best suggested solution for the problem that has been reported.

*2.1 Architecture Utilized for the Creation of the*

**A decision support system web-application for the management of forest road network**

*Web-based FRMP Tool*

In order to create this web application for internal usage, the MVC (Model-View-Controller) architecture has been chosen. Specifically, the ASP.NET by Microsoft with the "entity" framework is used. The MVC architectural pattern segregates the application into three main components: (a) the model, (b) the view, and (c) the controller (see Fig. 1 for a visual explanation of the basic MVC design pattern).

The ASP.NET MVC framework provides an alternative to the ASP.NET Web Forms pattern for creating Web applications. The ASP.NET MVC framework is a lightweight, highly testable presentation framework that (as with Web Forms-based applications) is integrated with existing ASP.NET features, such as master pages and membership-based authentication. The MVC framework is defined in the System.Web.Mvc assembly.

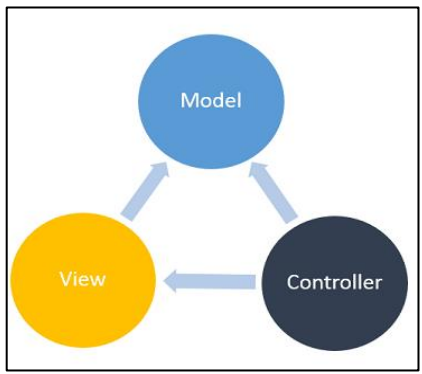

**Fig. 1   The basic MVC design pattern.**

MVC is a standard design pattern. The MVC framework utilized for the proposed web tool includes the following components:

• **Models:** Model objects are the parts of the application that implement the logic for the application's data domain. Often, model objects retrieve and store model state in a database. For example, a Product object might retrieve information from a database, operate on it, and then write updated information back to a Products table in a SQL Server database.

• **Views:** Views are the components that display the FRMP web application's UI (User Interface). This UI is created from the model data. An example would be an edit view of a Products table that displays text boxes, drop-down lists, and check boxes based on the current state of a Product object.

• **Controllers:** Controllers are the components that handle user interaction for the platform, work with the model, and ultimately select a view to render that displays UI. In the proposed MVC application, the view only displays information; the controller handles and responds to user input and interaction. For example, the controller handles query-string values, and passes these values to the model, which in turn might use these values to query the database.

The above described MVC pattern assists in creating applications that separate the different aspects of the application (input logic, business logic and UI logic), while providing a loose coupling between these elements. The pattern specifies where each type of logic should be located in the application. The UI logic belongs in the view, input logic belongs in the controller and business logic belongs in the model. This separation helps one manage complexity for building the web-based FRMP application, since it enables focusing on one aspect of the implementation at a time. For example, one may focus on the view without depending on the business logic.

There are certain advantages for the utilization of the MVC for the development of the FRMP Web-Based Application. First, the ASP.NET MVC framework makes it easier to manage complexity by dividing an application into the "model", the "view", and the "controller". Second, it does not use view state or server-based forms. This makes the MVC framework ideal for developers who want full control over the behavior of an application. It additionally uses a Front Controller pattern that processes Web application requests through a single controller. This enables one to design an application that supports a rich routing infrastructure. MVC also provides better support for TDD (Test-driven Development) ().

Finally, it performs well for web applications that are supported by large teams of developers and for web designers who need a high degree of control over the application behavior.

*2.2 Description of the FRMP Platform*

Hence, an internal-use application is designed, based on the previously described ASP.NET MVC framework, in the form of a web application consisting of two main parts. The first section describes the production of a land registry database that includes the process of registration and production of an integrated database that consists of road network data, such as the recording and classification of forest works and forest roads problems and their subsequent taxonomy in terms of the degree of their accessibility. In particular, the first section involves the registration and input of list of data relating to the potential problems of the forest road network of interest, such as collapses of rocks, landslides, ditch blockings, erosions, mobility of a road or a specific section of a road. The main benefit is that the existence of such a database can be used both for the better and the more objective and economical management and settlement of the forest roads. With the spatial information provided, there is a more prudent picture of the state of the forest road network of each forest cluster so that practical and reliable solutions can be adequately proposed and implemented.

The primary goal is to become as flexible and efficient as possible in the shortest possible time. By the electronic recording and creation of this database the scarce financial resources that one may have, can be prioritized and directed.

At a second stage (second part), the user (e.g. an administrator) of the web-application can observe the list of problems and reports, in combination with the forest map, so that the former can be effectively controlled. Hence, the administrator can choose one or multiple problems or reports massively to mark them as resolved, to make assignments for repairs and can calculate costs according to distance and type of the specific problem. More specifically, the collected data are stored into an SQL server. Subsequently it is desirable to be able to display them in a compact format e.g. through a browser to minimize the workload. It is desirable to have an interface which helps the "recorder" of the database to register a problem as fast as possible without making it difficult for the user. In addition, spatial data alongside the info will be served in order to make the user understand the location of the problem with simple directions from the caller (see flowchart in Appendix for a step by step description of the stages of FRMP application).

In order to store the data and to directly distribute it to the users of the application, a database that is 100% compatible with the automated system of the entity framework has been chosen. Hence, the database is an SQL Server 2014 Express edition.

For serving geo-spatial data the MapServer environment has been selected. MapServer is an open source development environment for building spatially enabled internet applications. It can run as a CGI program or via MapScript which supports several programming languages [using the simplified wrapper and Interface generator (SWIG)]. MapServer was developed by the University of Minnesota—thus, it is often and more specifically referred as "University of Minnesota, MapServer", to distinguish it from commercial "map server". MapServer was originally developed with support from NASA, which needed a way to make its satellite imagery available to the public. The basic source code for the replication of the web application is available upon request by the corresponding author.

Next, for a description of these two parts, a usage scenario will be presented. Let one suppose that the application is installed on a central server of a forestry service. Then it is divided into two main categories.

**A decision support system web-application for the management of forest road network**

For the purposes of this scenario it is assumed that a user of the application is someone who has the post of a phone operator in a forestry office and (s)he is called the "CCO (Call Center Operator)". Subsequently, a second user will be the manager of the application, called the "AM (Application Manager)". When the forestry office receives a call for reporting of a problem, at some particular geographical location of its forest road network, or a problem with a technical project, such as a ditch, then the CCO uses the application to record the problem through a simple menu (1st stage).

Next (2nd stage), the AM has the ability to view the complete list of problems and reports at the same time visually, for instance through a forest map so that the entire forest road network can be controlled. Specifically, the AM can choose one or more of the problems or reports, to mark them as resolved, assigning repairs, and being able to calculate costs according to distance and type of problem.

To illustrate the Forest Roads Management System in more detail, in the following a hypothetical scenario from a typical forest area located in Northern Greece is described. Fig. 2 shows the main screen of the application tool, for the "CCO" (stage 1).

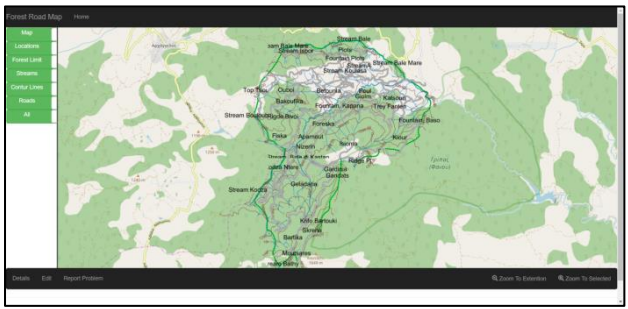

**Fig. 2 Forest road map as shown via the web application tool.**

With this simple interface, the ability to the user (e.g. the AM or CCO) to choose a part of the road to view in detail is provided. Let one suppose that one wants to concentrate on a specific part of the forest road network of the region of interest (Fig. 3). Upon choosing the part of the road network, the "CCO" person chooses a part and then has three options; "Details", "Edit" and "report a new problem" (Fig. 4).

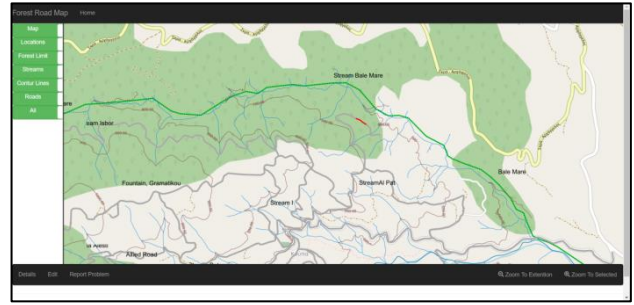

**Fig. 3 Selection of a specific part of the forest road map.**

If the "CCO" wants to register a new problem, he/she chooses the "report a new problem" tab. Then, (s)he logs the problem to the problems table by clicking on the map at the specific point or part of the road where the problem occurs. Subsequently, one chooses the type of the problem, types some description and presses "complete". By doing so the system registers the current time and date and the specific user that reported the problem.

By clicking on the "details" button, the CCO can get and accordingly edit all relative information of the specific part of the forest road in a simple format.

Next, from either the main menu or the "details" menu, the CCO can choose to edit the current information and insert updated information, through the "edit" tab menu of the web application (Fig. 5).

**A decision support system web-application for the management of forest road network**

Fig. 4 The "details" tab of the web application tool.

Fig. 5 The "edit road" tab menu of the web application.

After editing is completed, the user can return to the main initial screen of the web-application. If the user chooses to report a new problem (s)he can see a report problem form like the one presented in Fig. 6, titled as the "report problem" tab.

Fig. 6 The "report problem" tab menu of the web application.

By simply filling the associated tabs with all the relative information the register can subsequently save the created report. Later the "AM" can readily view the lists comprising of all problems associated with the forest road network of interest. Fig. 7 displays an overall report list comprising of all problems and issues in the particular part of the forest road of interest. From this menu, the AM can perform a variety of actions, such as edit, assign or close any open "Report list" in order to have the availability of the state of the forest roads always up to date. In particular, when the AM opens the main page of the web application and clicks on the part of the forest road network that is of interest, a menu appears with all the relevant information available, i.e. the road status (open/closed) and reported problems. Additional information includes road info such as: slope, shape, length etc. of the specific part of the forest road network (An alternative approach to the above described scenario is the option to give that access to the public so anyone can do exactly the same from his device of choice).

Fig. 7 The overall "report list" tab menu of the web application.

**A decision support system web-application for the management of forest road network**

When there is need for repairing the problematic parts of the forest road network, the commanding officer or the responsible employee of the forest office opens the commanding menu. (S)he chooses the types of problems (s)he wants to solve or the exact spot that needs to be repaired. Commanding officer checks all the info about the damage or problem and can calculates the costs with more accuracy. When the repair is complete one may mark the road as open again.

## 3. Results and Discussion

In the current section, an illustrative example of the utilization and the implementation of the proposed forest road network management web-tool is presented.

*3.1 Study Area*

The research area, namely the Koupa forest, is located in the Kilkis prefecture, of the Region of Central Macedonia, Northern Greece. Fig. 8 shows the visual presentation of the forest location at the geographical and political partition of the country, compared with the existing transport network.

The forest occupies the northeastern slopes of Mount Paiko. The area enclosed within the boundaries, using GIS map info was found to be 27,521.9 acres. The main forest species constituting the high forest are the oaks, i.e. species that can produce technical wood. The forest of Koupa presents a varied geomorphological relief. Its altitude varies between 380-1,540 m. There are many ridges across the forest in any direction, creating a variety of environments and orientations. Also, corresponding streams and tremors are often met in this forest area.

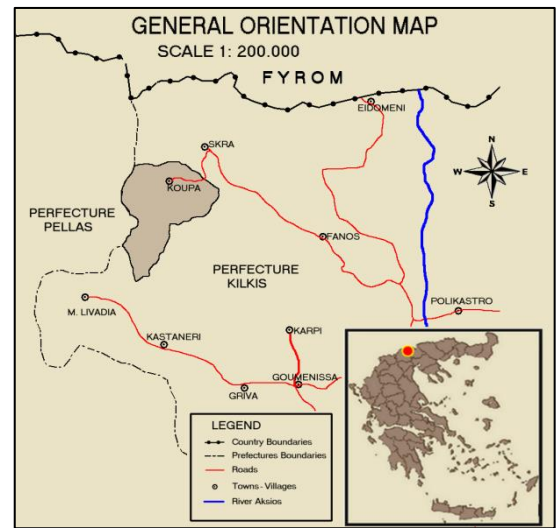

**Fig 8  Map of the study area**.

Note that as a study area a typical forestry area has been used, which due to certain characteristics is suitable for the current implementation. The total area covers 2,750 ha and the total length of the road network is 66 km. The existing road density is 24 m/ha, so a fully developed forest area with a rich forest road network with several intersections and alternate routes is utilized.

If one observes the climatic conditions of the particular forest area, it is found: The month of June has the lowest rainfall height of 25.3 millimeters. During winter (January to March), the rainfall amounts to 270.5 millimeters and the average monthly is 90.16 millimeters. During the blooming period from April to September, rainfall amounts to 381.2 millimeters and the average monthly is 63.53 millimeters. During the dry-heat period (June to September), the rainfall amounts to 205.5 millimeters and the average monthly is 51.37 millimeters.

The climatic environment that is observed is between the continental and the trans-European climate. The above is also evidenced by the fact that the beech appears in the forest at the altitude of 500 meters, indicating that the climatic conditions are favorable for its development. Snowfall is at a high level, while the northern exposure of most of the forest can lead to significant forest road problems.

**A decision support system web-application for the management of forest road network**

Taking into account the climate of the area in conjunction with the horticultural formation and the fact that most of the forest road network is a C-class forest roads, it is understood that after each winter period most of the forest road network needs important maintenance, rendering this forest area most suitable for the present illustration.

*3.2 Implementation of FRMP through Case Studies Scenarios*

In the current section, an implementation case scenario of the web-tool in order to show-case the strengths and benefits of the use of the FRMS program is presented.

Hence, in the following sub-section, a typical scenario of a person traveling to the forest of Koupa from say a point A to point F, covering a total distance of 7.769 kms, by using the shortest destination path, according to the road map is analyzed (Fig. 9). Then, it is further assumed that this person will be the driver of a truck carrying logging. Typically, the driver sees the map of the forest road network, chooses a path, fills the truck with gas and calculates the time he needs to load wood and transport it to the desired destination (Fig. 9 shows the shortest route, highlighted with the red line). This will be referred to as the A scenario.

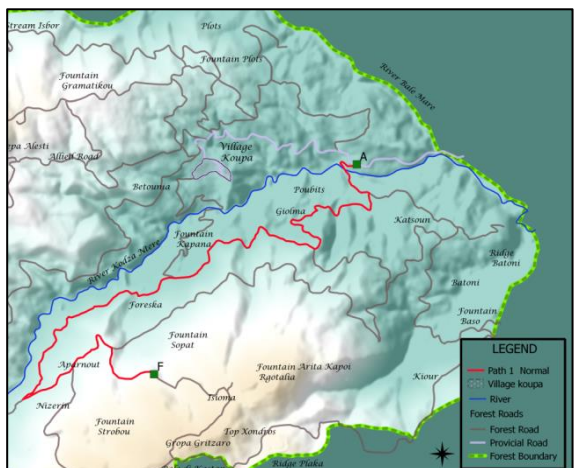

**Fig. 9 Shortest destination path according to scenario A.**

The corresponding estimated time $t$ for covering this distance is calculated based on the Greek fire service's official information, according to which a 10-tonne vehicle, moves at a speed of 30 km per hour at a forestry road [18]. For the example, half of that speed as the truck of the example can reach up to 30 tons of load and typically cannot approach the average speeds of fire brigade vehicles has been assumed. Hence, for the purposes of the present analysis it is assumed that a truck of 10-tonne covers the distance of 1 km in a typical forest road in approximately 4.3 minutes.

Next, let one suppose that the specific forest road network at the current time has a construction problem in the form of an obstacle, specifically at a certain point the accessibility is not possible (Fig. 10).

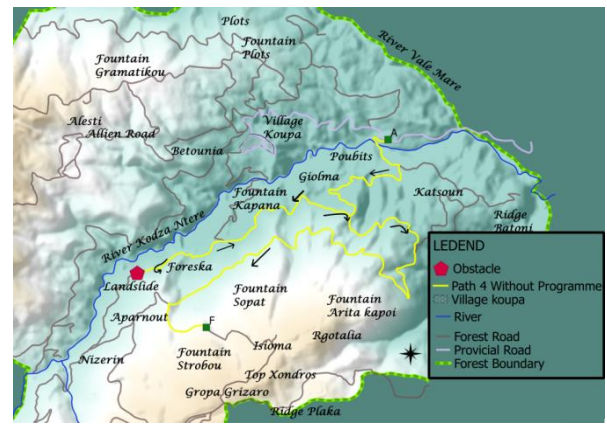

**Figure 10:** Destination path followed according to scenario B.

Without the use of the FRMS application, the driver chooses to follow a different route, after he confronts with the problem, hence essentially needs to turn back and choose an alternative route (this situation will be referred to as the B scenario). Then, the standard distance, time and costs will change depending on the next shortest path. Note also, that the alternative route may also end up with another problem which the driver is not aware of at the current time. However, for simplicity, let one assume that the alternative route does not suffer from additional problems. For the specific example, the driver will have to drive back 5,348 km to the nearest crossroads in order to take a longer route and drive for 6,107 additional kilometers.

Next, the two additional alternative scenarios that the driver can implement by using the FRMS decision-making program can be examined (Scenarios C and D). At the start of the travelling schedule, the driver checks the up-to-date web tool, hence will be informed on the road obstacle at the specific part of the route, and subsequently decide which is the shortest path. In the current situation, the program will inform the driver about the problem and suggest various alternative routes (Figs. 11 and 12 show two alternative routes as suggested by the FRMS). Also, the Forest Office would be able to provide more accurate info to the driver. Note also the possibility for the driver to check the status of the forest road network at a real-time upon the availability of internet access.

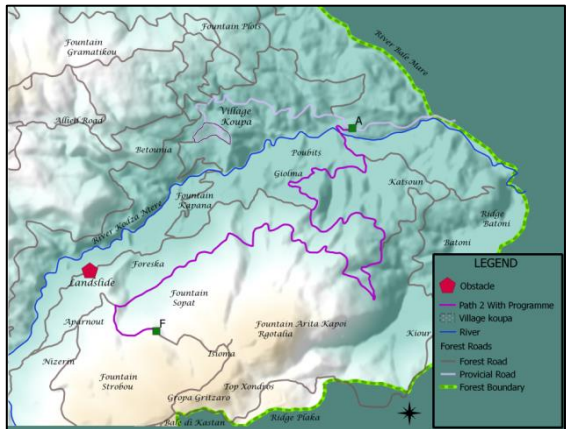

**Fig. 11 Destination path followed according to scenario C.**

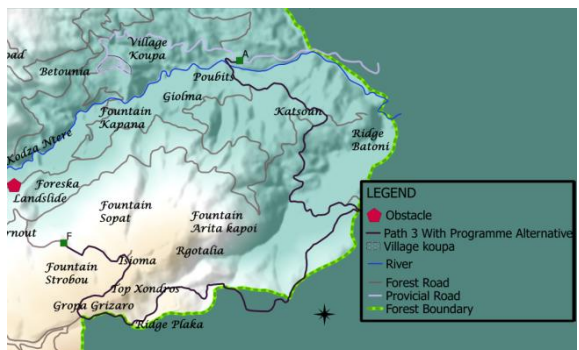

**Fig. 12 Destination path followed according to scenario D.**

The basic parameters of the four alternative scenarios (i.e. distance traveled and time consumed) are comparatively shown in Tables 1 and 2 for the alternative scenarios utilized. Specifically, Table 1 shows the values of the calculated parameters, whereas Table 2 presents the corresponding percentage of change of the parameters, with respect to the shortest available route that would have been followed by the driver in case of no road network issues (scenario A).

|  | Parameters | |
|---|---|---|
|  | Distance traveled ($d$) | Time ($t$) |
| Scenario A | 7,769 km | 33.29 min |
| Scenario B | 13,603 km | 58.29 min |
| Scenario C | 8,256 km | 35.38 min |
| Scenario D | 9,775 km | 41.89 min |

**Table 1 Calculated parameters' values for the alternative scenarios.**

|  | % of change in parameters in relation to scenario A | |
|---|---|---|
|  | % of change in $d$ | % of change in $t$ |
| Scenario B | 75.09% | 75.10% |
| Scenario C | 6.27% | 6.28% |
| Scenario D | 25.82% | 25.83% |

**Table 2 Calculated percentage of change in the parameters (with reference to scenario A).**

It is evident from the inspection of the results of

**A decision support system web-application for the management of forest road network**

Tables 1 and 2 that the possible utilization of FRMS would substantially reduce the total distance and time covered by the truck driver of the hypothetical scenario and corresponding costs, in case of a potential problem in the forest road network of the region. It is seen that without the assistance of FRMS the driver could potentially double the total distance covered (approximately 75% increase in distance covered), whereas under the FRMS application the corresponding percentages are between approximately 6.3% and 25.8%. Accordingly, the improvement of additional time consumed for bypassing the hypothetical obstacle through the use of the web-tool is varying between 68.82% and 49.27% for the scenarios C and D, respectively.

As a result, using the FRMS not only helps the Forest Office cut back the costs of repairing the damages of the forest facilities like stuck drainage pipes but can also take a major role in the information of the transports of all kinds through the forest roads.

Overall, from the previous analysis, results and examples, it is seen that using the proposed computer application results in considerable improvements in forest management parameters, such as distance covered and time consumed, enhancing thus the optimal utilization of any forest road network.

## 4. Conclusions

The main objective of the current study was the proposal and development of a decision making web-application that aims to assist in the management of forest road networks in a holistic way. The proposed application provides an easy to use and useful tool that combines geoprocessing and geospatial technologies for the efficient optimization of the management of a forest road network. With the spatial information provided, there is a more prudent picture of the state of the forest road network of each forest complex. In this way, practical and reliable solutions can be proposed and found, so that by imprinting them on three-dimensional ground models, they will help in making more reliable decisions.

Overall, the proposed forest road management web-based tool has been shown to give satisfactory results, by the optimization of both resources used and time consumed. Nevertheless, the adoption of this forest road network management model has been proven to be a useful decision making tool, although it needs further improvements. Finally, among the strengths of the proposed tool, one may prioritize the fact that the former is a light application, meaning that the end point user does not have to have a computer that cost much. It is costless since it is an open source application in all of the aspects of development and it can be modified to work with any open source type of database or program and finally it can be used as a PPGIS (Public Participation GIS) system to enrich the data gathering from other sources rather than only emergency calls. As regards potential weaknesses, these include the fact that certain aspects of the application are designed to be carried out manually instead of being fully automatic. In addition, no sensors or cameras are currently available for auto feedback. However, these limitations are the subject of ongoing research towards the improvement of the current web application.

**A decision support system web-application for the management of forest road network**

**A decision support system web-application for the management of forest road network**

## Appendix

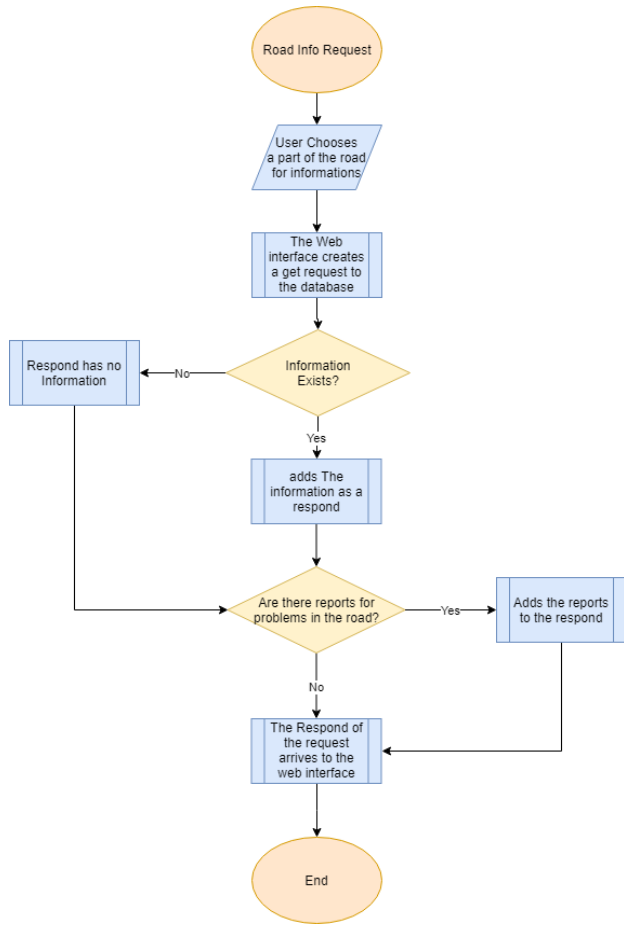

Flowchart of theseparate steps of FRMP. application.